\documentclass[preprint,amsmath,amssymb,aps,showkeys,showpacs]{revtex4}
\usepackage[colorlinks=true, pdfstartview=FitV, linkcolor=blue, citecolor=red, urlcolor=magenta]{hyperref}
\usepackage{graphicx}
\usepackage{latexsym}
\usepackage{amsmath}
\usepackage{amsfonts}
\usepackage{amssymb}
\usepackage{verbatim}
\usepackage{dcolumn}
\usepackage{amssymb}
\usepackage{bm}
\usepackage{float}
\usepackage{subfigure}
\usepackage[english]{babel}
\newcommand{\be}{\begin{equation}}
\newcommand{\ee}{\end{equation}}
\newcommand{\bea}{\begin{eqnarray}}
\newcommand{\eea}{\end{eqnarray}}

\begin{document}

\newcommand{\NITK}{
\affiliation{Department of Physics, National Institute of Technology Karnataka, Surathkal  575 025, India}
}

\title{Coexistent Physics and Microstructure of the Regular Bardeen Black Hole in Anti-de Sitter Spacetime}
\author{Ahmed Rizwan C.L.}
\email{ahmedrizwancl@gmail.com}
\NITK
\author{Naveena Kumara A.}
\email{naviphysics@gmail.com}
\NITK
\author{Kartheek Hegde}
\email{hegde.kartheek@gmail.com}
\NITK
\author{Deepak Vaid}
\email{dvaid79@gmail.com}
\NITK

\begin{abstract}
We study the phase structure and the microscopic interactions in regular Bardeen AdS black hole. The stable and metastable phases in the black hole are analysed through coexistence and spinodal curves. The solutions are obtained numerically as the analytic solution to the coexistence curve is not feasible. The $P_r-T_r$ coexistence equation is obtained using a fitting formula. The coexistence and spinodal curves are plotted in $P_r-T_r$ and $T_r-V_r$ planes to explore the phase structure of the black hole. In the second part of our study, we were able to probe the microscopic interactions of regular Bardeen AdS black hole using the novel Ruppeiner geometry proposed by S.W. Wei \emph{et.al} Phys. Rev. Lett.123, 071103 (2019). It is found that the microscopic interactions are not same in the small black hole (SBH) and large black hole (LBH) phases. In the SBH phase, there exists a repulsive interaction in the microstructure in the low temperature regime. In contrast, the microstructure associated with the LBH phase has attractive interaction throughout the parameter space. We found that, along the coexistence temperature both the SBH and LBH branches diverge to negative infinity with a critical exponent equal to $1/2$. 
\end{abstract}


\maketitle


\section{Introduction}
Black hole thermodynamics is a subject that helps a theoretical physicist to unveil the deep connection between gravitation, quantum theory and statistical physics. Even after 50 years of Bekenstein and Hawking's work, the problem of black hole entropy and the temperature is not well understood \cite{Hawking:1974sw, Bekenstein1972, Bekenstein1973, Bardeen1973}. During the past three decades, in quest to find a quantum gravity theory the focus has directed towards the black holes in asymptotically anti-de Sitter(AdS) spacetimes. It is mainly due to the pioneering work of Hawking and Page, which explored the phase transition between the radiation and a large black hole \cite{Hawking:1982dh}. Black holes in AdS cavity provided necessary thermal stability to this thermodynamic system. But the Smarr relation for AdS black is found inconsistent with the first law \cite{Kastor:2009wy}. To rectify this problem, cosmological constant $\Lambda$ was considered as a thermodynamical variable. In the first law, it was interpreted as the thermodynamic pressure, and its conjugate quantity was found to be the geometrical volume. This has extended the first law of black hole thermodynamics with a necessary $VdP$ term  \cite{Kastor:2009wy, Dolan:2011xt}. In this extended phase space, thermodynamics of charged AdS black hole was found to be analogous to a van der Waals fluid system \cite{Kubiznak2012, Gunasekaran2012, Kubiznak:2016qmn}. This has led to a new arena in the black hole physics called the \emph{black hole chemistry}.

  This macroscopic picture of the black hole is used to propose a phenomenological model for black hole microstructure \cite{Ruppeinerb2008}. Even though the microscopic information is not a requirement for the thermodynamics, it may be used for quantum gravity studies. Prominent model for drawing microscopic information from thermodynamics is the Ruppeiners's thermodynamic geometry \cite{Ruppeiner95}. It is constructed on the equilibrium state space in the context of thermodynamic fluctuation theory, but can be useful in studying black holes too \cite{Ruppeinerb2008}. Through gaussian fluctuation moments, a Riemannian geometry is constructed in the thermodynamic equilibrium space, whose metric tells us about the fluctuations between the states. This method is applied to van der Waals fluids and to a variety of other statistical systems \cite{Ruppeiner95, Janyszek_1990, Oshima_1999x, Mirza2008, PhysRevE.88.032123}. These studies show that the thermodynamic geometry encodes the information about the microscopic interaction. The thermodynamic scalar curvature $R$ is proportional to the correlation volume of the underlying system. The sign of $R$ indicates the type of interaction in the microstructure, positive for repulsive and negative for attractive interactions. In recent times, there has been a lot of interest in the thermodynamic geometry to investigate critical phenomenon and microstructure of various black holes in AdS spacetime \cite{Wei2015, Sahay:2016kex, Guo2019, Miao2017, Zangeneh2017, Wei:2019ctz, Kumara:2019xgt, Kumara:2020mvo, Xu:2019nnp, Chabab2018, Deng2017, Miao2019a, Chen2019, Du2019, Dehyadegari2017, Ghosh:2019pwy, Ghosh:2020kba}.

Recently, a novel approach for Ruppeiner geometry was developed to explore the missing information due to the singularity in the scalar curvature \cite{Wei2019a}. This is mainly due to the vanishing of heat capacity at constant volume. The new normalised scalar curvature takes care of this problem. A metric can be defined by Taylor expanding the Boltzmann entropy around the equilibrium value. The thermodynamical coordinates were chosen to be the temperature and volume, and the Helmholtz free energy was chosen as the thermodynamic potential. Applying this method to the van der Waals (vdW) fluid,  it was found that dominant interaction in the microstructure is attractive throughout the parameter space. Utilizing the analogy with vdW fluid, thermodynamic geometry of a charged AdS black hole is analysed. In contrast to the vdW fluid, the interaction is not attractive over the entire parameter space. Even though the interaction is attractive for the large black hole (LBH) every where and small black hole (SBH) for most of the parameter space, there exists a weak repulsive interaction in the SBH phase at very low temperatures \cite{Wei2019a, Wei2019b}. Interestingly, this behaviour is not universal for all asymptotically AdS black holes. In the case of five-dimensional neutral Gauss-Bonnet black hole, interaction similar to vdW fluid is observed, with a dominant attractive interaction throughout the SBH and LBH phases \cite{Wei:2019ctz}. Soon later, work is extended to 4-dimensional Gauss-Bonnet black holes \cite{Wei2020}. Subsequently, the microscopic interactions for $4-D$ AdS topological black holes dRGT massive gravity were studied\cite{Yerra:2020oph, Wu:2020fij}. Microstructure was found to be distinct, with the presence of both repulsive and attractive interactions in both the SBH and LBH phases. In our recent paper, we have investigated the microstructure of regular Hayward and Born-Infeld AdS black holes \cite{Kumara:2020ucr, NaveenaKumara:2020biu}. The microscopic interactions observed is similar to the case of charged AdS black holes in regular Hayward case. Where as, Born-Infeld AdS black holes show a reentrant phase transition, which has a distinct microstructure. Apart from these studies, the study of microstructure using this novel method is limited to a few black holes. Motivated by the recent progress, here we explore the phase structure and microstructure of a regular Bardeen AdS black hole.

    Regular black holes are the ones which does not possess a singularity at the centre. Even though it is in the domain of quantum gravity theory to obtain a singularity free solution, a phenomenological model can be constructed in the classical gravity. Firstly such a regular solution was derived by Bardeen \cite{Bardeen1973}. Later many have found that regular black holes can be an exact solution to gravity coupled with a non-linear electromagnetic source \cite{AyonBeato:1998ub, AyonBeato:2000zs, Hayward:2005gi}. We have studied phase transitions and thermodynamic geometry of regular black holes in our recent papers \cite{Rizwan2019, Rajaniheat, Naveen2019photon}. It is noticed that the presence of magnetic monopole charge imparts a  phase structure to the regular black holes similar to the electric charge. So we find it interesting to probe the microstructure corresponding to the magnetically charged Bardeen black holes in asymptotically AdS spacetimes.

The paper is organised as follows. In section \ref{metric}, we review the action and derivation of the regular Bardeen black hole in AdS spacetime. In section \ref{Phase structure}, we mainly focus on the thermodynamics and phase structure of the black hole. Then the Ruppeiner geometry and analysis of critical features are discussed in section \ref{TG}. The final section \ref{summary} is dedicated for the summary and conclusions.


\section{Regular Bardeen AdS Black hole}
\label{metric}
The Bardeen black hole emerges as the solution to the Einstein's gravity coupled to a non-linear electrodynamics source with a negative cosmological constant $\Lambda$. We will consider an action,
\begin{equation}
\mathcal{S}=\frac{1}{16\pi}\int d^4x \sqrt{-\tilde{g}}\left(R-2\Lambda-\mathcal{L}(\mathcal{F})\right),\label{Action}
\end{equation}
where $R$ denotes the Ricci scalar, $\tilde{g}$ the determinant of metric tensor $\tilde{g}_{\mu\nu}$, and $\Lambda$ is the cosmological constant. $\mathcal{L}(\mathcal{F})$ is the Lagrangian density of non-linear electrodynamics, which is the function of the field strength $\mathcal{F}=F_{\mu\nu}F^{\mu\nu}$ with $F_{\mu\nu}=\partial_\mu A_\nu- \partial_\nu A_\mu$. Variation of the action (\ref{Action}) leads to Einstein's and Maxwell's equations of motion, given by
\begin{equation}
G_{\mu\nu}+\Lambda g_{\mu\nu}= T_{\mu\nu},\quad\quad \nabla_\mu \left(\frac{\partial\mathcal{L}(\mathcal{F})}{\partial\mathcal{F}}F^{\mu\nu}\right)=0 \quad\text{and}\quad \nabla_\mu\left(*F^{\nu\mu}\right)=0.\label{eqns}
\end{equation}
$G_{\mu\nu}$ is the Einstein tensor and $T_{\mu\nu} = 2\left(\frac{\partial\mathcal{L}(\mathcal{F})}{\partial\mathcal{F}}F_{\mu\lambda}F^{\lambda}_\nu-\frac{1}{4}g_{\mu\nu}\mathcal{L}(\mathcal{F})\right)$ is the energy-momentum tensor. The Lagrangian density in the case of Bardeen black holes is,
\begin{equation}
\mathcal{L}(\mathcal{F})= \frac{12}{\alpha}\left(\frac{\sqrt{\alpha\mathcal{F}}}{1+\sqrt{\alpha \mathcal{F}}}\right)^{5/2},
\end{equation}
where $\alpha$ is a positive quantity with a dimension $[\text{Length}]^2$. We take the following ansatz for Maxwell's field tensor,
\begin{equation}
F_{\mu\nu}=2 \delta^\theta_{[\mu}\delta^\phi_{\nu]}Q(r)\sin \theta.
\end{equation}
But from Maxwell's equations (\ref{eqns}), $dF=\frac{dQ(r)}{dr}dr\wedge d\theta\wedge d\phi=0$ which require $Q(r)$ to be a constant $Q_m$. For a spherically symmetric solution, the non-vanishing components of Maxwell's field tensor are $F_{tr}$ and $F_{\theta\phi}$. Since we are interested in a magnetically charged regular solution, we choose gauge potential and Maxwell's field tensor to be,
\begin{equation}
A_\mu= Q_m \cos\theta \delta^\phi_\mu,\quad F_{\theta\phi}=-F_{\phi\theta} = Q_m \sin\theta,
\end{equation}
where $Q_m$ is the magnetic monopole charge. The scalar function $F$ is obtained from $F_{\theta\phi}$ as,
\begin{equation}
F=\frac{2Q_m^2}{r^4}.
\end{equation}
We can rewrite Lagrangian density $\mathcal{L}(\mathcal{F})$ as a function of radial distance,
\begin{equation}
\mathcal{L}(r)=\frac{12}{\alpha}\left(\frac{2\alpha Q_m^2}{r^2+2\alpha Q_m^2}\right)^{5/2}.
\end{equation}
A static spherically symmetric solution for the Einstein's equation can be put in the form,
\begin{equation}
ds^2=-f(r)dt^2+\frac{dr^2}{f(r)}+r^2\left(d\theta^2+\sin^2 \theta d\phi^2\right),
\end{equation} 
with the metric function $f(r)=1-\frac{2 m(r)}{r}-\frac{\Lambda r^2}{3}$.  Making use of the line element, the Einstein's equation is solved for fixing the functional form of $m(r)$. $G_{tt}$ and $G_{rr}$ components of Einstein's equation read as,
\begin{align}
\frac{1}{r^2}\partial_r m(r)-\Lambda &=\frac{1}{4}\mathcal{L}(r),\\
\frac{1}{r}\partial_r^2 m(r)-\Lambda&=\left(\frac{1}{4}\mathcal{L}(r)-
\frac{\partial\mathcal{L}}{\partial\mathcal{F}} F_{\theta\phi}F^{\theta\phi}\right).
\end{align}
Integrating the above differential equations, we obtain the mass function $m(r)$ for a regular Bardeen AdS black hole as,
\begin{equation}
m(r)=\frac{\Lambda r^3}{6}+ \frac{Mr^3}{\left(g^2+r^2\right)^{3/2}},
\end{equation}
where $M$ is the mass of the black hole and  $g$ is the charge parameter related to total charge $Q_m$,
\begin{equation}
Q_m=\frac{g^2}{\sqrt{2\alpha}}.
\end{equation}
So, the line element for Bardeen-AdS black hole is written with the metric function, 
\begin{equation}
f(r)=1-\frac{2 M r^2}{\left(g^2+r^2\right)^{3/2}}-\frac{\Lambda r^2}{3}.
\end{equation}
\section{Thermodynamics and phase structure}
\label{Phase structure}
In this section, we review thermodynamics of the black hole in an extended phase space, where the cosmological constant $\Lambda$ is  given the status of a dynamical variable pressure $P$. It can be justified from Smarr relation and first law of black hole thermodynamics in the asymptotically AdS spacetimes. The thermodynamic pressure $P$ is related to $\Lambda$ as,
\begin{equation}
P=-\frac{\Lambda}{8\pi}.\label{pressure}
\end{equation}  
Firstly, we write the first law of black hole thermodynamics and Smarr relation for the magnetically charged Bardeen AdS black hole \cite{Fan:2016hvf, Fan:2016rih},
\begin{align}
dM=&TdS+\Psi dQ_m+VdP+\Pi d\alpha,\\
M=&2(TS-VP+\Pi \alpha)+\Psi Q_m.
\end{align}This can be obtained either from Komar integral, or from scaling argument presented in the paper by Kastor \emph{et al } \cite{Kastor:2009wy}. Notice that there exists additional terms $\alpha$ and $\Pi$, they are the parameters related to the non-linear electromagnetic field and its conjugate potential respectively.
We can write black hole mass $M$ using the condition $f(r_h)=0$  at the event horizon $r=r_h$,  
\begin{equation}
M=\frac{\left(g^2+r_h^2\right)^{3/2} \left(8 \pi  P r_h^2+3\right)}{6 r_h^2}.\label{Mass}
\end{equation}
 The Hawking temperature of the black hole is obtained as,
\begin{equation}
T=\left.\frac{f'(r)}{4\pi}\right|_{r=r_h}=-\frac{g^2}{2 \pi r (g^2 + r_h^2)} + \frac{r_h}{4 \pi \left(g^2 + r_h^2\right)} + \frac{2 P r_h^3}{g^2 + r_h^2}, \label{temperature}
\end{equation}
where we have used equations (\ref{Mass}) and (\ref{pressure}) for mass $M$ and pressure $P$. The required thermodynamic quantities for our analysis, volume $V$ and entropy $S$ can be obtained from the first law, 
\begin{align}
V=&\left(\frac{\partial M}{\partial P}\right)_{S,Q_m}=\frac{4}{3} \pi  \left(g^2+r_h^2\right)^{3/2},\\ 
S=&\int \frac{dM}{T}= \pi r_h^2 \left. \left[\left(1-\frac{2g^2}{r_h^2}\right) \sqrt{1+\frac{g^2}{r_h^2}}+\frac{3 g^2}{r_h^2} \log \left(\sqrt{g^2+r_h^2}+r_h\right)\right]\right.\label{Volume}.
\end{align}
The thermodynamic stability of black hole is specified by the heat capacity at constant pressure $C_P$ and volume $C_V$, which is determined as,
\begin{align}
C_P=&T\left(\frac{\partial S}{\partial T}\right)_P=\frac{2 S \left(\pi  \beta^2+S\right) \left(-2 \pi  \beta^2+8 P S^2+S\right)}{2 \pi ^2 \beta^4+\pi  \beta^2 S (24 P S+7)+S^2 (8 P S-1)},\\
C_V=&T\left( \frac{\partial S}{\partial T}\right)_V=0.
\label{cv}
\end{align}
One can obtain the equation of state, $P=P(V,T)$, utilising the expression for the Hawking temperature (\ref{temperature}) and thermodynamic volume (\ref{Volume}),
\begin{equation}
P=\frac{\left(\frac{6 V}{\pi }\right)^{2/3} \left(-1+2 \pi  T \sqrt{\left(\frac{6 V}{\pi }\right)^{2/3}-4 g^2}\right)+12 g^2}{2 \pi  \left(\left(\frac{6 V}{\pi }\right)^{2/3}-4 g^2\right)^2}.
\end{equation}
We study the phase structure of the black hole in the canonical ensemble with a fixed monopole charge $g$. The $P-V$ isotherms show a first-order van der Waals fluid like phase transition between two phases, namely, the \emph{small black hole} (SBH) and the \emph{large black hole} (LBH) phase. The critical point is obtained from the inflection point of the $P-V$ isotherm,
\begin{equation}
\left.\frac{\partial P}{\partial V}\right|_{r=r_h} = \left.\frac{\partial^2 P}{\partial V^2}\right|_{r=r_h}=0.
\end{equation}  
The critical quantities temperature $(T_c)$, pressure $(P_c)$ and volume $(V_c)$, thus obtained are given below, 
\begin{align}
T_c = &\frac{\left(17-\sqrt{273}\right) \sqrt{\frac{1}{2} \left(\sqrt{273}+15\right)}}{24 \pi  g},\\
P_c = &\frac{\sqrt{273}+27}{12 \left(\sqrt{273}+15\right)^2 \pi  g^2},\\
V_c = &\frac{4}{3} \pi g^3 \left(1+\frac{1}{2} \left(\sqrt{273}+15\right)\right)^{3/2}.
\end{align}
Using critical quantities, we define the following reduced coordinates,
\begin{equation}
T_r=\frac{T}{T_c},\quad P_r=\frac{P}{P_c},\quad V_r=\frac{V}{V_c}.
\end{equation}
We can rewrite the equation of state in the reduced parameter space as,
\begin{equation}
P_r=\frac{\left(\sqrt{273}+15\right)^2 {V_r}^{2/3} \left(3 \left(\sqrt{273}+17\right)-4 T_r \sqrt{\sqrt{273}+15} \sqrt{-2+\left(\sqrt{273}+17\right) {V_r}^{2/3}}-\frac{18}{{V_r}^{2/3}}\right)}{\left(\sqrt{273}+27\right) \left(2-\left(\sqrt{273}+17\right) {V_r}^{2/3}\right)^2}.
\end{equation}
The phase transition in a Bardeen AdS black hole was extensively studied by AG Tzikas \cite{Tzikas:2018cvs}. But there are couple of things left out in the analysis due to the difficulty in inverting the equation of state and solving $r_h$ as function of pressure and temperature, $r_h=r_h(P,T)$. We address this problem numerically and obtain the coexistence equation from the swallow tail behaviour of Gibbs free energy.

We begin our analysis from the Gibbs free energy, which is defined as the Legendre transform of enthalpy, recall that addition of $VdP$ term leads to the identification of mass as enthalpy $H$. The Gibbs free energy reads as,
\begin{equation}
G(P,r_h,g)= H-TS. \label{Gibbseqn}
\end{equation}
And change in the Gibbs free energy,
\begin{equation}
dG =-SdT+VdP+\Phi dg.
\end{equation}
\begin{figure}[H]
\centering
\subfigure[ref1][\, $ G_r \quad \text{vs} \quad T_r$]{\includegraphics[scale=0.8]{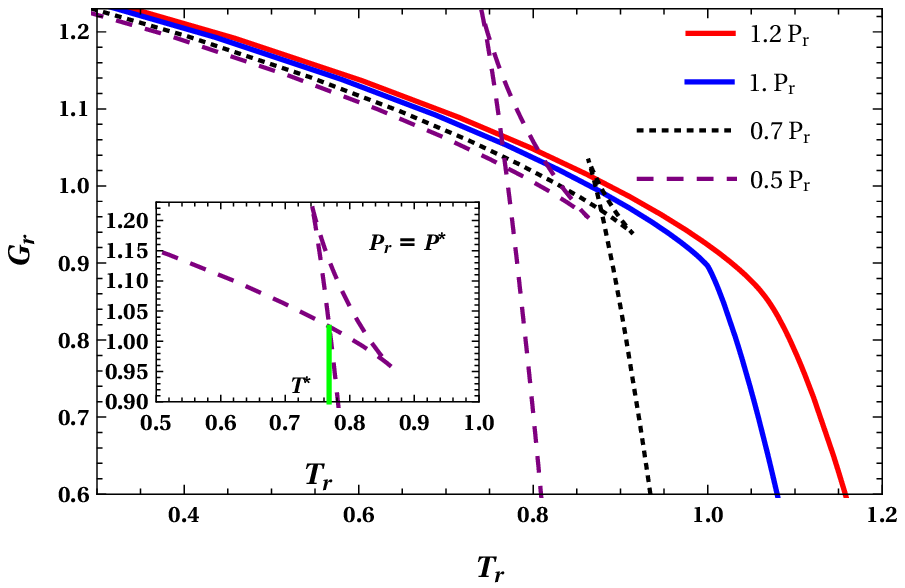}
\label{BGibbs}}
\qquad
\subfigure[ref2][\, $P_r \quad \text{vs} \quad V_r$]{\includegraphics[scale=0.8]{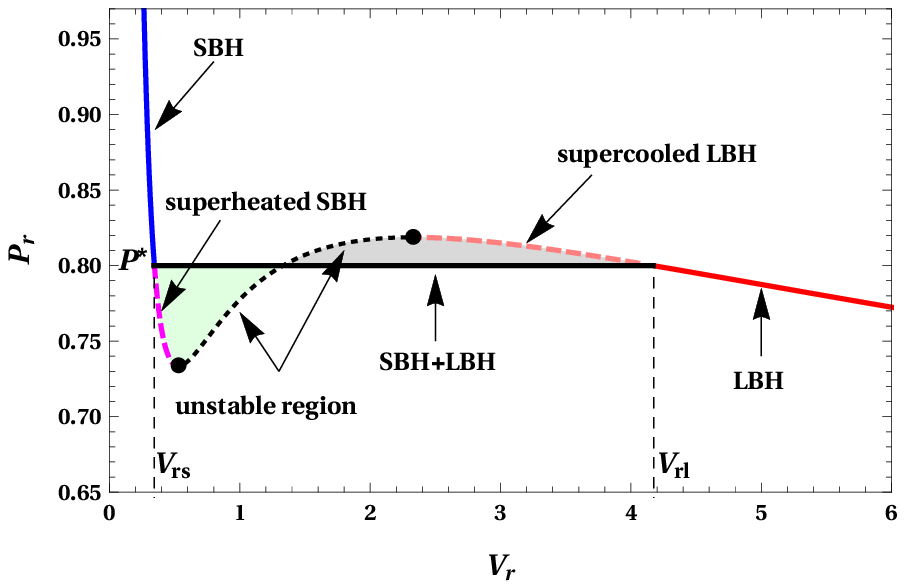}
\label{BPV}}
\caption{In fig \ref{BGibbs}, the Gibbs free energy $G_r$ is plotted as a function of reduced temperature $T_r$ for different reduced pressure $P_r$. The swallow tail behaviour is exhibited when $P_r<1$. In inlets, a magnified view of the swallow tail at a pressure $P_r=P^*<1$ is shown. In fig \ref{BPV} Isotherm with reduced temperature $T_r=T^*<1$ under Maxwell's construction is shown. The SBH, superheated SBH, unstable region, supercooled LBH and stable LBH phases are labelled in the figure \ref{BPV}. }
\end{figure}
The Gibbs energy and its change is important in determining thermodynamic stability of a system. In an equilibrium state, when the pressure, temperature and charge are fixed, $G$ takes a minimal value. But often writing $G$ explicitly as the function of temperature and pressure, $G(P,T)$ is difficult. We can obtain the Gibbs free energy plots parametrically using equations (\ref{Gibbseqn}) and (\ref{temperature}). In fig \ref{BGibbs}, we plot reduced Gibbs free energy $(G_r)$  as the function of reduced temperature $(T_r)$ for different pressures. When reduced pressure $P_r<1$, we can see a  \textquotedblleft swallow tail behaviour \textquotedblright which is a typical signature of a first-order phase transition. A close observation will reveal that there are three regions in the swallowtail, two branches corresponding to the stable SBH and LBH phases, and a tail connecting these two. As the difference in Gibbs free energy between two branches becomes zero, the transition takes place between SBH and LBH phases.  At the critical pressure or below these two branches become distinct. But they do intersect at certain temperature $T^*$, where two phases coexist. We use these data to plot the coexistence curve and fit into a coexistence equation numerically. Using the fitting method, we obtain an expression for the coexistence equation,
\begin{align}
P_r=&-0.022 + 0.625 T_r -6.726 T_r^2 + 48.312 {T_r}^3 - 
 194.636 {T_r}^4 + 511.332 {T_r}^5\\
    & - 887.151 {T_r}^6 + 1009{T_r}^7 -  722.605 {T_r}^8 + 295.313 {T_r}^9 - 52.4384 {T_r}^{10},\\
 T_r \in & (0,1).   
\end{align}
\begin{figure}[H]
\centering
\subfigure[ref1][]{\includegraphics[scale=0.8]{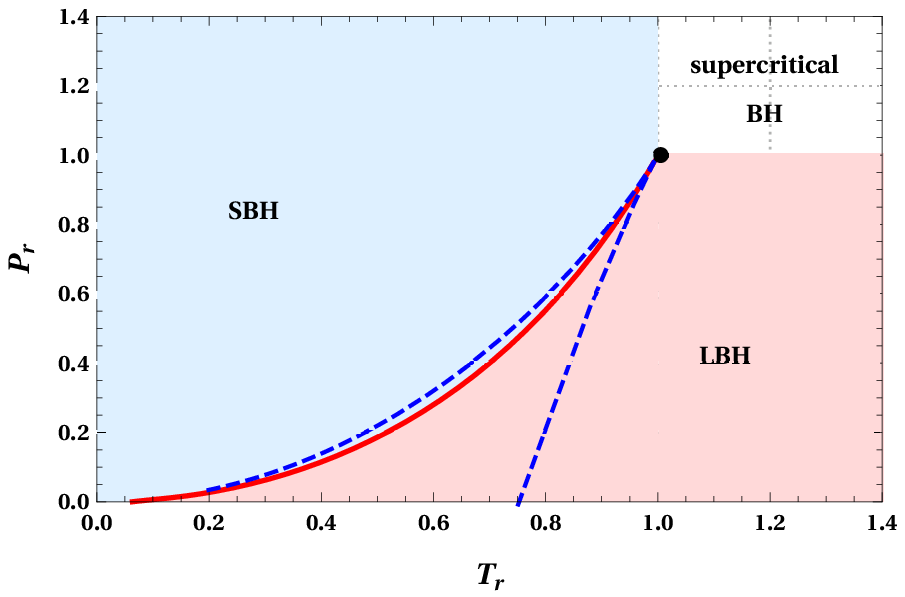}
\label{BPT}}
\qquad
\subfigure[ref2][]{\includegraphics[scale=0.8]{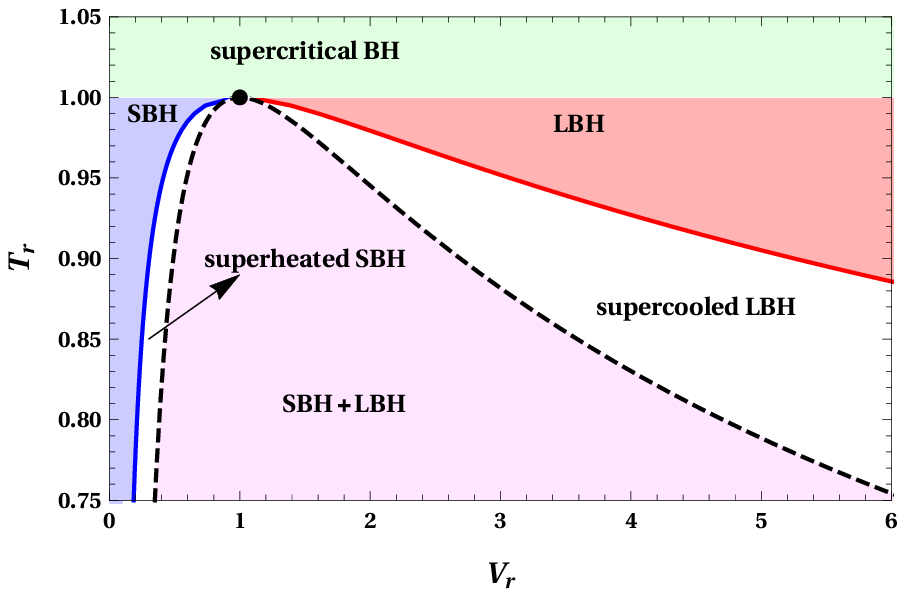}
\label{BVT}}
\caption{Coexistence and spinodal curves in $P_r-T_r$ and $T_r-V_r$ plane. The coexistence curve is shown in solid line and the spinodal curves are shown in dashed line.  }
\label{fig1}
\end{figure}
In Fig \ref{BPT}, we have obtained $P_r-T_r$ coexistence diagram using the fitting formula. The red line is the locus of coexistence phase pressure and temperature $(P^*,T^*)$.  The light magenta shaded region below the curve is the the LBH phase and the light green color region depicts SBH phase. The black point at the coordinate $(1,1)$ denotes critical point and above that (unshaded region) differentiation of phases is impossible known as supercritical region. In the fig \ref{BPT}, along the coexistence curve, we have also shown the spinodal curve marked by the blue dashed line. It is plotted using the condition,
\begin{equation}
\left( \partial _{V_r} T_r \right)_{P_r}=0,\quad
\text{or}\quad \left( \partial _{V_r} P_r \right)_{T_r}=0.
\end{equation}
The spinodal curve equation obtained from above condition is of the form,
\begin{align}
T_{rsp}=&\frac{3 \left(17+\sqrt{273}\right) \left(-10+\left(17+\sqrt{273}\right) {V_r}^{2/3}\right) \sqrt{-2+\left(17+\sqrt{273}\right){V_r}^{2/3}}}{4 \sqrt{15+\sqrt{273}} \left(-4+\left(17+\sqrt{273}\right) {V_r}^{2/3}+\left(17 \sqrt{273}+281\right) {V_r}^{4/3}\right)},
\\
P_{rsp}=&\frac{3 \left(15+\sqrt{273}\right)^2 \left(-12 {V_r}^{2/3}+9 \left(17+\sqrt{273}\right) {V_r}^{4/3}-\left(281+17 \sqrt{273}\right) {V_r}^2\right)}{2 \left(27+\sqrt{273}\right) {V_r}^{2/3} \left(3 \left(17+\sqrt{273}\right) {V_r}^{2/3}-\left(\left(285 \sqrt{273}+4709\right) {V_r}^2+4\right)\right)}.
\end{align}

\begin{figure}[H]
\centering
\subfigure[ref3][\, $\log\left(\Delta V_r\right) \quad \text{vs} \quad T_r$]{\includegraphics[scale=0.8]{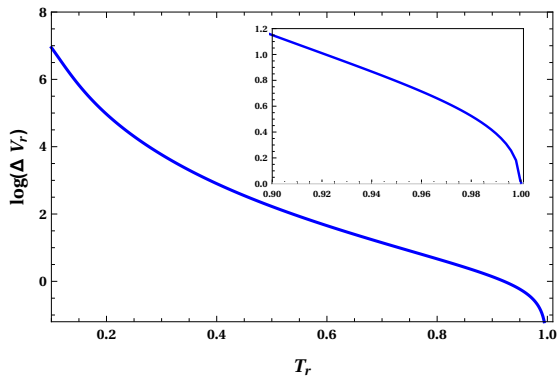}
\label{BVTorder}}
\caption{The volume change $\Delta V_r=V_{l}-V_{s}$ as a function of reduced temperature $T_r$. Magnified view near the critical point is shown in the inlets. }
\end{figure}

Using parametric plots, we have obtained spinodal curves, which is the locus of extreme points separating metastable SBH and LBH phases from the unstable region. And as it is evident from figures \ref{BPT} and \ref{BVT}, the spinodal curve is the envelope of the saturated mixture of SBH and LBH phases. The spinodal curve also has a maximum at the critical point. The coexistence phase structure is shown in $T_r-V_r$ plane along with the spinodal curve in fig \ref{BVT}. Careful analysis of curves will find that there are five regions, namely SBH phase, LBH phase, supercritical phase, metastable superheated SBH and supercooled LBH phase. In both the plots in $P_r-T_r$ and $T_r-V_r$ planes, the extremal point coincides with the extrema in the spinodal curves.  To have more clarity, we can turn to $P_r-V_r$ isotherms. When the reduced temperature $T_r$ of isotherm is below 1, we can see an oscillating behaviour with an inflection point at the critical point. At any temperature $T^*$, corresponding to coexistence $T_r-V_r$ curve which is obviously bounded below 1, the isotherm consists of fore mentioned five regions. In fig \ref{BPV}, we have presented  labelled $P_r-V_r$ plot indicating different regions. For Maxwell's construction, a vertical line is drawn at the pressure $P^*$. The line divides the isotherm into two equally occupied regions satisfying Maxwell's equal-area law. At the temperature $T^*$, the volume of the SBH and LBH phases is $V_s$ and $V_l$ respectively. The $V_s$ and $V_l$ are obtained from $T_r-V_r$ coexistence curve. The terminology used here is defined parallel to the analogous van der Waals fluid system. From the $P_r-V_r$ isotherm, we see that SBH phase (thick blue) can exist till the pressure $P^*$ where it has the volume $V_s$. When the pressure is reduced below $P^*$, the system moves to a superheating phase without undergoing a transition. This phase denoted by the pink dashed portion is the superheated SBH phase. This state is metastable in the sense it can undergo a phase transition with even small fluctuation. End of this metastable phase is marked by a black dot which represents the spinodal curve. Further, there exists a small unstable region with positive slope denoted by the black dotted line in the figure \ref{BPV}. This unstable region terminates at the extremum, from there system  moves to another metastable state known as supercooled LBH. The unstable region is separated from the metastable region by the spinodal curve. The supercooled LBH phase is marked as the magenta dashed line in the plot. The system continues in this state till $P^*$, after that system, undergoes rapid expansion with a slight change in pressure. The volume acts as an order parameter during this transition. At the critical point, the difference between the volumes of SBH and LBH phases vanishes and they form a single supercritical phase. These regions are also portrayed in the coexistence curve in $T_r-V_r$ plane fig \ref{BVT}. Using the numerical method, we plot the volume change $\Delta V_r=V_l-V_s$ as the function of reduced temperature $T_r$ in fig \ref{BVTorder}. It shows that $\Delta V_r$ approaches zero at the critical point and monotonously increases when temperature is reduced. The series expansion of $\Delta V_r$ around the critical point reads,
\begin{equation}
\Delta V_r = 4.03691 (1-T_r)^{0.537073}.
\end{equation}
 The critical exponent is $0.537$, which is approximately equal to  the universal value $1/2$. This result is similar to the one earlier obtained in different black holes \cite{Wei2019a,Wei2019b,Wei:2019ctz,Kumara:2020ucr,Wu:2020fij}.


\section{Microstructure of the Bardeen AdS Black Hole}
\label{TG}
It is known from early works of George Ruppeiner, that the information about the thermodynamic phase transition is captured in the thermodynamic geometry constructed in the thermodynamic parameter space $(P,T,V,S)$. In this section, we study the critical behaviour and the microstructure of the Bardeen black hole using the novel Ruppeiner geometry put forward by Wei \emph{et.al} \cite{Wei2019a}, where  $T$ and $V$ are chosen as the fluctuation coordinates. The line element is written in $(T, V)$ coordinates as,
\begin{equation}
dl^2=\frac{C_V}{T^2}dT^2-\frac{\left(  \partial _V P\right)_T }{T}dV^2.
\label{line}
\end{equation}
And the normalised scalar curvature $R_N$ is obtained from the above line element as,
\begin{equation}
R_N= R C_V= \frac{\left(\partial_V P\right)^2-T^2 \left(\partial_{V,T}P\right)^2+2T^2\left(\partial_V\right) \left(\partial_{V,T,T}P\right)}{2\left(\partial_V P\right)^2}.\label{RN}
\end{equation}
We have seen from equation (\ref{cv}) that the heat capacity ($C_V$) at constant volume vanishes for the black hole. This can result in a singularity and this singular behaviour in the curvature scalar $R$ is rectified by multiplying it with $C_V$. 

\begin{figure}[H]
\centering
\subfigure[ref1][]{\includegraphics[scale=0.8]{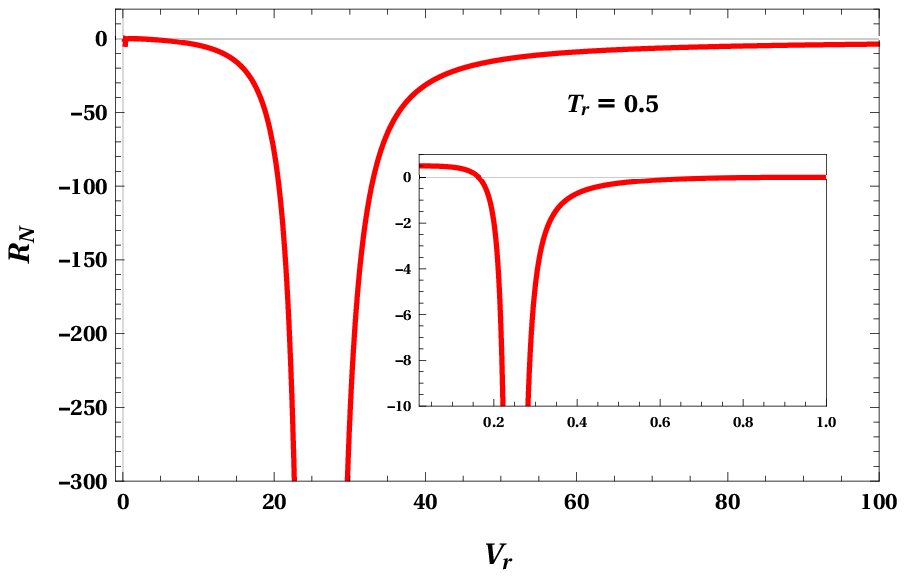}\label{BRNV1}}
\qquad
\subfigure[ref2][]{\includegraphics[scale=0.8]{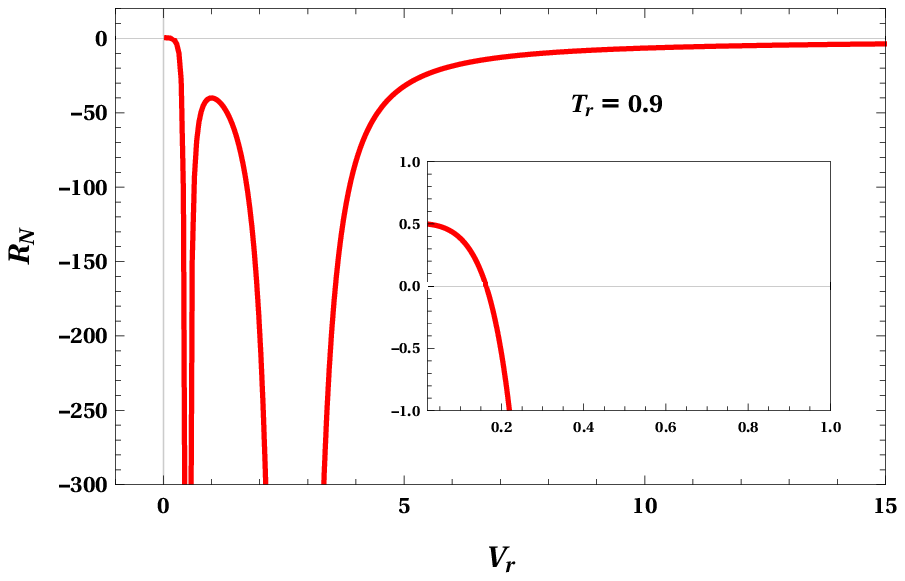}\label{BRNV2}}
\subfigure[ref1][]{\includegraphics[scale=0.8]{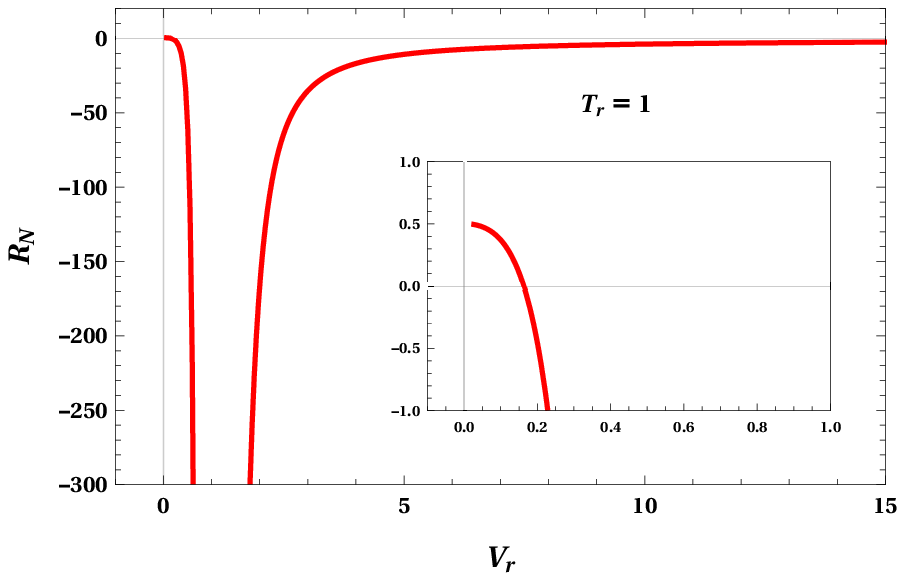}\label{BRNV3}}
\qquad
\subfigure[ref1][]{\includegraphics[scale=0.8]{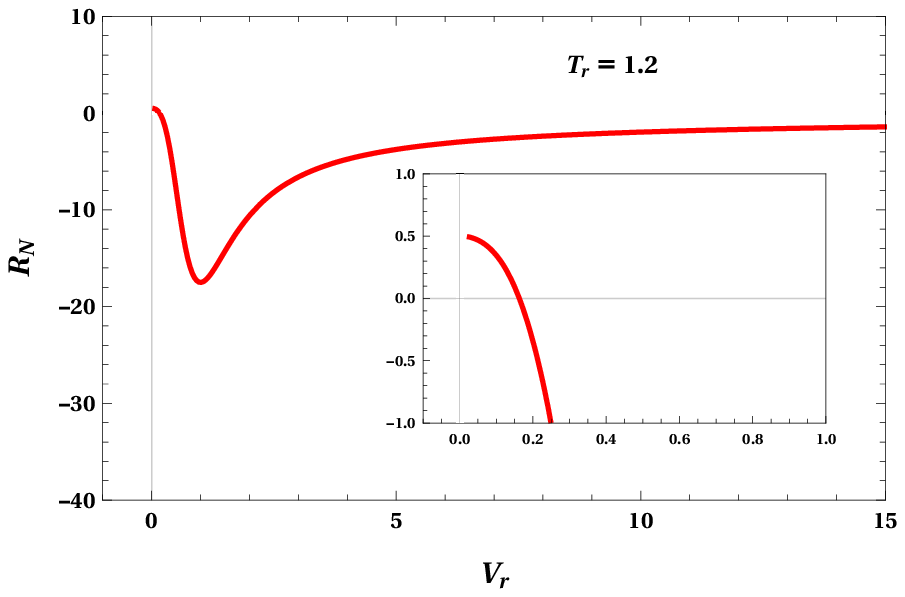}\label{BRNV4}}
\caption{The normalised curvature scalar $R_N$ is plotted against the reduced volume $V_r$ at different reduced temperature $T_r$. }
\label{RNplots}
\end{figure}

\begin{figure}[H]
\centering
\subfigure[ref1][]{\includegraphics[scale=0.8]{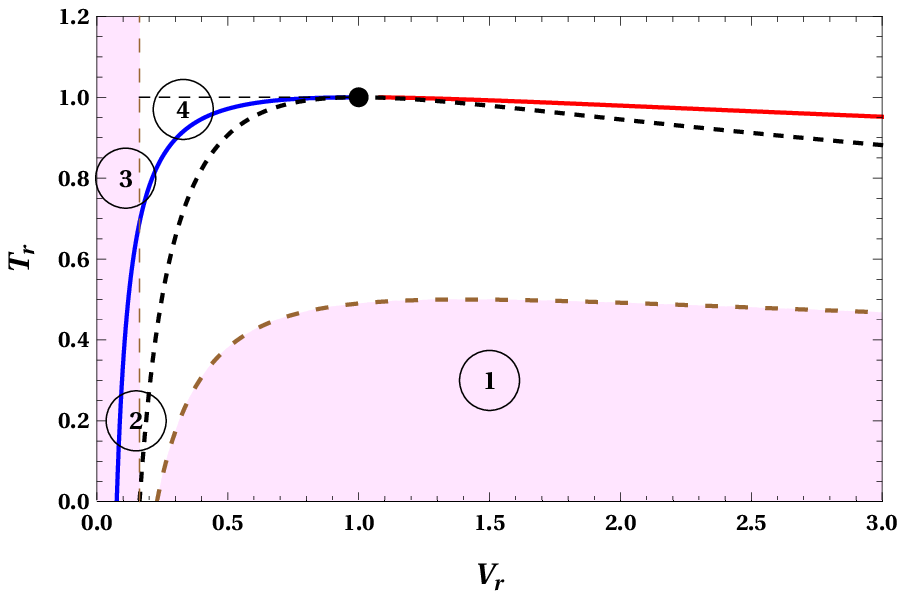}\label{BSignCoexSpin}}
\qquad
\subfigure[ref2][]{\includegraphics[scale=0.8]{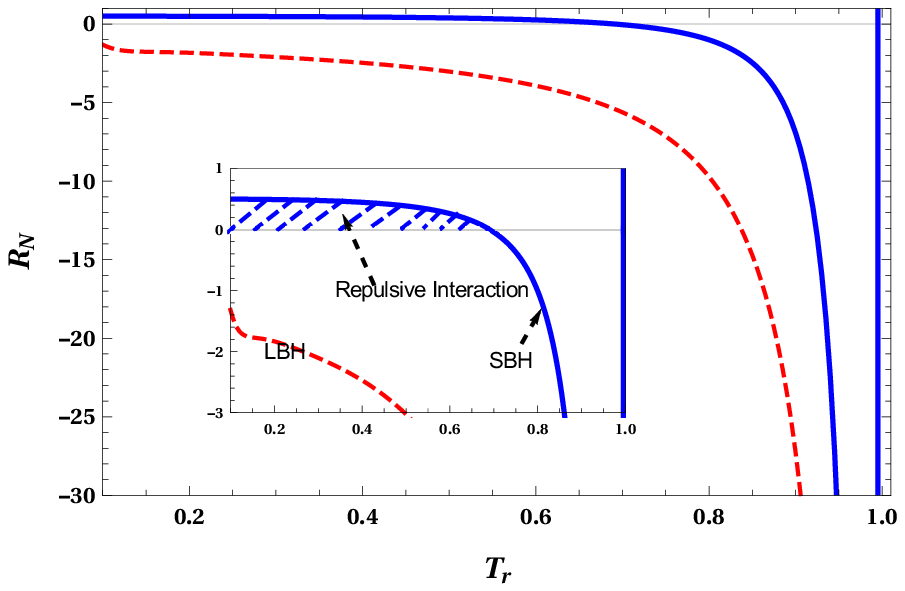}\label{BRT}}
\caption{ \ref{BSignCoexSpin}: The sign-changing curve of $R_N$ along with the coexistence and spinodal curves. \ref{BRT}: The behaviour of normalised curvature scalar $R_N$ along the coexistence line. The red (solid) line and blue (dashed) line correspond to a large black hole and a small black hole, respectively. The inlet shows the region where the SBH branch takes a positive $R_N$ value. }
\end{figure}

  The normalised scalar $R_N$ gives the information about the microscopic interactions present in the black hole. The metric tensor is calculated using the line element (\ref{line}) and $R_N$ is obtained from the equation (\ref{RN}). $R_N$  hence obtained is a complicated expression $R_N(T,V,g)$. After converting it in the reduced coordinates $R_N(T_r,V_r)$, it is plotted against the reduced volume $V_r$ with a fixed temperature is shown in fig (\ref{RN}). In reduced coordinates, $R_N$ is independent of monopole charge $g$. From the figures \ref{BRNV1} and \ref{BRNV2}, we can see that $R_N$ has two divergent points below the critical point $T_r<1$. And when the temperature becomes equal to the critical temperature $T_r=1$, these divergences merge and shoot up at the point $V_r=1$ as showed in fig \ref{BRNV3}. As expected, the divergence vanishes for all temperatures above the critical temperature $T_r>1$ fig \ref{BRNV4}.  This shows that the information about the phase transition and critical phenomenon is well expressed by the normalised curvature scalar $R_N$. Interestingly, these two divergent points correspond to the metastable points in the spinodal curves. We can notice that even though $R_N$ is negative for most of the parameter space, there exists a small range where it is positive which are shown in the inlets of Fig (\ref{RNplots}). The sign changing curve is plotted in the $T_r-V_r$ plane utilising the condition scalar curvature $R_N=0$.
 The scalar curvature $R_N$ vanishes and changes its sign at the point $T_0$, given by
\begin{equation}
T_0=\frac{T_{rsp}}{2}=\frac{3 \left(\sqrt{273}+17\right) \left(\left(\sqrt{273}+17\right) {V_r}^{2/3}-10\right) \sqrt{\left(\sqrt{273}+17\right) {V_r}^{2/3}-2}}{8 \sqrt{\sqrt{273}+15} \left(\left(\sqrt{273}+17\right) {V_r}^{2/3}+\left(17 \sqrt{273}+281\right) {V_r}^{4/3}-4\right)}.
\end{equation}
 Sign change also happens at the point,
 \begin{equation}
V_r=V_0=\frac{5}{8} \sqrt{\frac{5}{2} \left(4709-285 \sqrt{273}\right)}.
\end{equation}
The sign-changing curve distinguishes regions of negative $R_N$ from the positive. As we know the scalar curvature $R_N$ tells about the microscopic interaction. Positive $R_N$ means a repulsive interaction and negative $R_N$ signifies an attractive interactions in the microstructure. To have more clarity, we have placed all three plots, coexistence, spinodal and sign-changing plots in a single plot fig \ref{BSignCoexSpin}. Different regions in the figure \ref{BSignCoexSpin} correspond to the stable and metastable phases. The light magenta shaded region under the sign changing curve has positive and unshaded region has negative $R_N$ . The region \textcircled{1}, area common between the spinodal curve and sign-changing curve is a saturated SBH+ LBH phase. This phase always has a repulsive interaction in the microstructure with a positive $R_N$. Next, the left-most region bounded inside the line $V_r=V_0$, also has a repulsive interaction with a positive $R_N$. In that, region \textcircled{3} is an SBH phase and region \textcircled{2} is a metastable superheated SBH phase. Nevertheless, there is a stable SBH portion which lies outside this positive $R_N$ region marked with \textcircled{4}, has a attractive interaction in the microstructure. All other phases, supercooled LBH as well as stable LBH phase have negative $R_N$, with dominant attractive interaction. It affirms that, there exists attractive and repulsive interaction in the black hole microstructure, a hint of this is observed in $R-V_r$ plot (\ref{RNplots}).
 In Fig \ref{BRT}, normalised curvature scalar $R_N$ is plotted as a function of temperature along the coexistence line. This is obtained numerically from the  $P_r-T_r$ coexistence fitting equation. Both of the branches, SBH and LBH , diverge to infinity at the critical temperature $T_r=1$. Besides this, we can see that sign of $R_N$ is always negative for the LBH phase, but the same is not true for the SBH phase. As shown in inlets of fig \ref{BRT}, in the small temperature range, there exists a region with positive $R_N$. Even though the scalar curvature decreases with temperature for both the branches, LBH branch never attains positive $R_N$. Like in the previous plot (\ref{RNplots}), this also leads to the conclusion that there exists attractive and repulsive interaction in the microstructure of SBH phase. We also notice that the intensity of attractive interaction in the SBH phase is stronger than that of the LBH phase. It can happen due to the strong correlation between the black hole molecules in the SBH phase than in loosely correlated LBH phase. This behaviour is similar to van der Waals liquid-gas system, where the attractive interaction in the liquid phase is more intense than in the gaseous phase.

 \begin{figure}[H]
\centering
\subfigure[ref1][]{\includegraphics[scale=0.8]{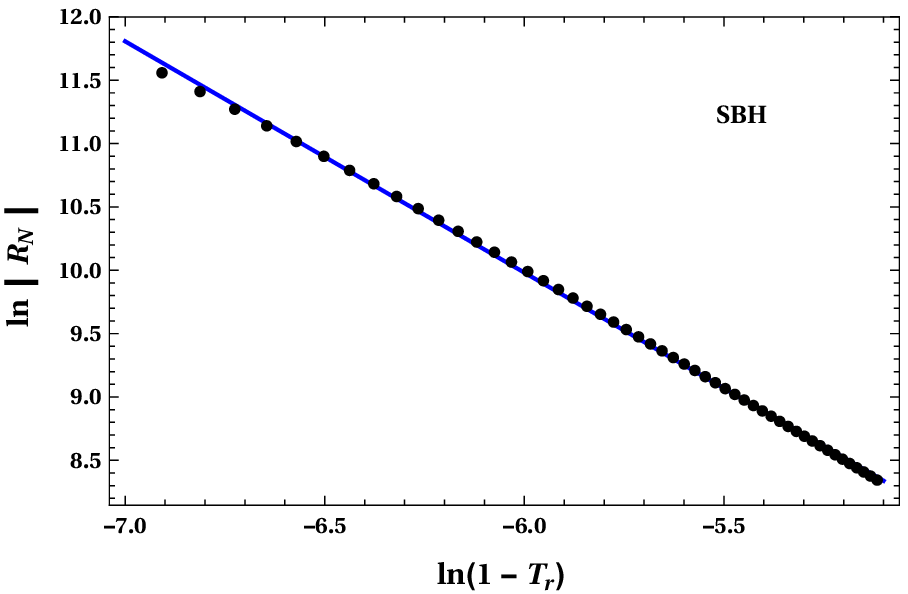}\label{RTfitSBH}\label{SBH}}
\qquad
\subfigure[ref2][]{\includegraphics[scale=0.8]{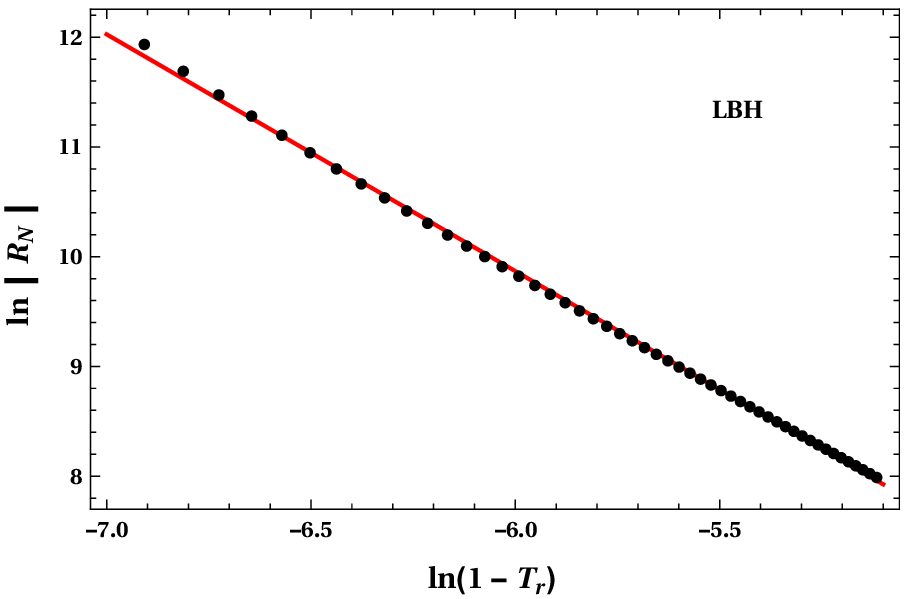}\label{RTfitLBH}\label{LBH}}
\caption{ The behaviour of  scalar curvature $\ln|R_N|$ near the critical point is shown in terms of $\ln(1-T_r)$ for the case of LBH in fig \ref{LBH} and SBH in fig \ref{SBH}. The numerical data points are marked by black dotes and line obtained from fitting formula are in solid blue line for SBH and in red line for LBH phase. }
\end{figure}
Finally, we can find the critical exponent corresponding to the divergence of $R_N$ along the coexistence line for the SBH and LBH branches. This can be obtained numerically assuming that $R_N$ has the form,
\begin{equation}
R_N\sim\left(1-T_r\right)^p.
\end{equation}
Taking logarithm on both sides, it reduces to,
\begin{equation}
\ln|R_N|=-p\ln\left(1-T_r\right)+q.
\end{equation}
We have numerically generated data for $R_N$ as a function of coexistence temperature $T_r$ in the range $0.9$ to $0.999$. Along the SBH branch, we can fit the data as,
\begin{equation}
\ln|R|=-1.82528\ln\left(1-T_r\right)-0.970181 .\label{RLtSBH}
\end{equation}
Similarly, for LBH branch, we obtain,
\begin{equation}
\ln|R|=- 2.15789\ln\left(1-T_r\right)-3.07915 . \label{RLtLBH}
\end{equation}
We have plotted these equations separately for SBH and LBH phases in the figures \ref{SBH} and \ref{LBH} along with the numerical data points. These plots show a great consistency in the solid lines and numerical data. Apart from the numerical errors, the results show that the critical exponent $p$ is approximately equal to 2. From equations (\ref{RLtSBH}) and (\ref{RLtLBH}), we can write,
\begin{equation}
R_N\left(1-T_r\right)^2=-\exp^{-\left(1.82528+2.15789\right)/2}=-0.132038.
\end{equation}
This ratio is slightly higher than the universal ratio $-1/8$ found in vdW system and other AdS black holes. Taking the numerical errors into account, the result we obtained is very close to the universal ratio.

\section{Summary and Conclusions}
\label{summary}

In this paper, we have concentrated mainly on studying the thermodynamics and microstructure of regular Bardeen AdS black holes. Information about the coexistence phases missing in earlier studies in the literature are addressed. We have dedicated initial sections for obtaining coexistence $P_r-T_r$ equations from the Gibbs free energy plots. The Gibbs free energy in reduced coordinates is plotted as a function of reduced temperature $T_r$ with a fixed pressure $P_r$. The appearance of swallowtail behaviour in these plots below the critical pressure is used to generate data for obtaining the coexistence equation. As it is difficult to obtain coexistence equation analytically in Bardeen black holes, from Maxwell's equal-area law or Gibbs free energy, we have used the fitting formula. Through the coexistence equation, different regions in $P_r-V_r$ isotherm is analysed at a reduced temperature $T_r<1$. It is noticed that a first-order phase transition analogous to vdW system takes place between stable SBH and LBH phases. Besides, there exists metatable superheated SBH and supercooled LBH phases. The stable and metastable phases are distinguished from each other by plotting a spinodal curve. The unstable regions are removed through the Maxwell's constructions. These distinct phases in the black hole are studied through the coexistence and spinodal curves in $P_r-T_r$ and $T_r-V_r$ planes. The change in volume $\Delta V_r=V_{rl}-V_{rs}$ acts as an order parameter during the SBH-LBH phase transition. Near the critical point, the critical exponent is calculated, which matches with the universal value of $1/2$.  

  In the second half of the paper, we have studied the thermodynamics through Ruppeiner geometry. The novel method proposed by Wei\emph{et.al} is used to calculate the thermodynamic scalar curvature \cite{Wei2019a}. Using the reduced equation of state for Bardeen black hole, novel scalar curvature is calculated and plotted against reduced volume $V_r$ at different temperatures. The critical behaviour is well captured in the plots with appearance and disappearance of divergences below and above the critical point. Moreover, it is noticed that scalar curvature attains both positive and negative values in the plots. The sign of $R_N$ encodes the information about the microscopic interactions. This leads one to an inference that both attractive and repulsive interaction exists in the black hole microstructure. To have more details on the microstructure, we have analysed the behaviour of scalar curvature along the coexistence curve. In the absence of analytical expression for the coexistence curve, we depend on the numerical methods for obtaining $R_N$ vs $T_r$ plots for SBH and LBH branches. Both the branches diverge to negative infinity at critical point $T_r=1$. Except for the divergence, the microstructure of the SBH and LBH are distinct. The LBH phase always has a larger $|R_N|$ than the SBH branch. Moreover, SBH branch attains positive $R_N$ in the small temperature range. This is in agreement with $R_N-V_r$ plots, SBH microstructure has repulsive as well as attractive interactions, but LBH microstructure has only attractive interactions. This is in contrast to the vdW fluid system where only attractive dominant interactions are present between the molecules. Our results imply that the phase transition leads to a change in the microstructure of the regular
   Bardeen AdS black holes. A similar type of behaviour is observed in charged  AdS black hole and regular Hayward AdS black hole. But this feature is not universal, in five-dimensional neutral Gauss-Bonnet black hole case, only attractive interaction present in the entire parameter space similar to van der Waals fluid. In Born-Infeld AdS black holes and massive gravity theories, the nature of interaction depends on the value of coupling and massive parameter respectively.


\acknowledgments
Authors A.R.C.L., N.K.A. and K.H. would like to thank U.G.C.Govt. of India for financial assistance under UGC-NET-SRF scheme.


  \bibliography{BibTex}

\end{document}